# Time-domain Channel Property Feedback in 5G-Advanced and Beyond

Per Ernström, Siva D. Muruganathan, Keerthi Kumar Nagalapur, Fredrik Athley

Ericsson

*Abstract*—**The availability of time-variability information of a channel between a network node and a user equipment at the network side allows a 5G network to optimally configure its parameters to maximize both user and system performance. In the Release 18 enhancements of 5G-Advanced, a time-domain channel property (TDCP) feedback that indicates the degree of time-variability of the channel is being standardized. In this article, we describe the standardized TDCP feedback and its applications. The benefit of the feedback is further illustrated through numerical evaluations.**

## I. INTRODUCTION

5G NR is very flexible with its support for different subcarrier spacings; different demodulation reference signal (DMRS), downlink (DL) channel state information reference signal (CSI-RS), and uplink (UL) sounding reference signal (SRS) configurations; and different types of channel state information (CSI) feedback [1]. A great flexibility implies that a network is challenged with the task of configuring parameters to maximize both user and system performance. The network can, to some extent, use the information contained in feedback received from user equipment (UE) or explicit channel measurements to determine the most suitable parameters.

Time-variability of a channel between the network and a UE, i.e., a measure of how fast the channel is varying with time, is a key property that would allow the network to configure parameters to optimize performance. Knowledge about how fast the channel is varying with time could be used for multiple use-cases. We list some important use-cases below.

1) Switching between CSI Type-I and Type-II based precoding: multi-user (MU) multiple input multiple output (MIMO) transmission using feedback based on Type-I codebook suffers in performance due to coarse CSI. Type-II codebook enables feedback of a richer CSI which can be used to maximize the spectral efficiency [3]. However, the fine CSI feedback received using Type-II codebook can get outdated quickly and degrade the MU-MIMO throughput performance when the channel varies faster than a tolerable threshold. The availability of channel time-variability information to the network would allow it to switch from Type-II codebook to Type-I codebook when the channel time-variability is higher than a threshold.

2) DL DMRS time-direction density configuration: NR supports a flexible configuration of 0 to 3 additional DMRS symbols, equivalent to 1 to 4 DMRS symbols, distributed along the time-direction in a slot. A higher number of DMRS in a slot allows a more accurate channel estimation in time-varying channels at the cost of a higher DMRS overhead. The availability of channel time-variability information to the network would allow it to configure an appropriate DMRS density that enables sufficiently accurate channel estimation with the least overhead.

3) Switching between different CSI-RS periodicities: CSI-RS can be transmitted periodically from a network node to enable the UEs to estimate the channel. The channel estimates can be further used by the UEs to compute and feedback a CSI report. If a network node is aware of the time-variability of channels to the UEs, it can configure the periodicity of CSI-RS to enable channel estimation at a sufficient time separation with the least overhead.

4) Switching between different CSI feedback periodicities: Due to a much smaller total transmit power in UL compared to DL, the spectral efficiency in UL is typically much smaller than in DL. Therefore, it is essential to minimize the overhead in UL transmissions. When a network is aware of the time-variability of the channels, it can configure or trigger the UEs to feedback CSI reports in UL with the largest possible periodicity that ensures in-time feedback to reduce the UL feedback overhead.

5) Switching between different UL SRS periodicities: Similar to the CSI-RS transmission periodicity, if a network node is aware of the time-variability of channels to the UEs it can configure the UEs to transmit SRS at a periodicity that is sufficient for in-time channel estimation and has the least overhead.

Per Ernström is with Ericsson Standards and Technologies, Stockholm, Sweden (e-mail: per.ernstrom@stosn.com).

Siva D. Muruganathan is with Ericsson Standards and Technologies, Ottawa, Canada (e-mail: siva.muruganathan@ericsson.com).

Keerthi Kumar Nagalapur is with Ericsson Research, Gothenburg, Sweden (e-mail: keerthi.kumar.nagalapur@ericsson.com).

Fredrik Athley is with Ericsson Research, Gothenburg, Sweden (e-mail: fredrik.athley@ericsson.com).

6) Switching between reciprocity- and feedback-based precoding: due to a much smaller total transmit power in UL compared to DL (~20 dB smaller), precoding based on reciprocity is typically reliable only for users with sufficient UL signal-to-noise ratio (SNR) and feedback-based precoding is necessary for users with low UL SNRs. Along with the UL SNR information, the network can use the channel time-variability information to switch between a reciprocity and a feedback-based precoding scheme while considering the overhead resulting from different DL CSI-RS, UL SRS, and CSI feedback periodicities.

7) Switching between different reciprocity-based precoding schemes: for users with sufficient UL SNR where reciprocity-based precoding is used, some schemes are more robust to channel time-variations than others. For example, schemes based on grid-of-beams precoding are more robust to channel time-variations compared to schemes that use singular vectors of the measured channel. A measure of channel time-variation can therefore help the network to choose the type of the reciprocity scheme.

Channel variability could in principle be estimated based on the UL SRS or physical uplink shared channel (PUSCH) DMRS. As noted before, total transmit power in UL is much smaller compared to DL. To get sufficient accuracy, however, the signal needs to be wideband. A wideband DMRS cannot be ensured since it depends on PUSCH resource allocation; and a wideband UL SRS would consume a large fraction of available UL resources and may suffer from low SNR due to the lower total transmit power constraint. It is therefore beneficial to let the UEs perform the channel measurement based on the CSI-RS-for-tracking (TRS) and report the channel time-variability to the network node. Moreover, in a typical deployment, TRS are periodically transmitted to enable the UEs to perform fine time-frequency synchronization and estimate parameters such as Doppler spread that are necessary to process physical downlink shared channel (PDSCH), and UEs could reuse TRS for estimating channel variability.

The time-domain channel property (TDCP) feedback feature is being standardized by 3GPP for 5G NR Release 18. The TDCP report is being designed with the objective of feeding back information related to time-variability of the channel to the network. In the following sections, we describe the TDCP feature. We also show that it provides good performance gains through numerical evaluations of two use-cases: a) switching between CSI Type-I and Type-II precoding and b) DMRS time-direction density configuration. The switching between CSI Type-I and Type-II precoding use-case is very challenging since the switching point between CSI Type-I and Type-II precoding is at low time-variability of channels. A small change in the channel is harder to estimate than a large change. If the TDCP feedback performs well for this use-case, one can therefore expect it to perform well also for the other use-cases described above.

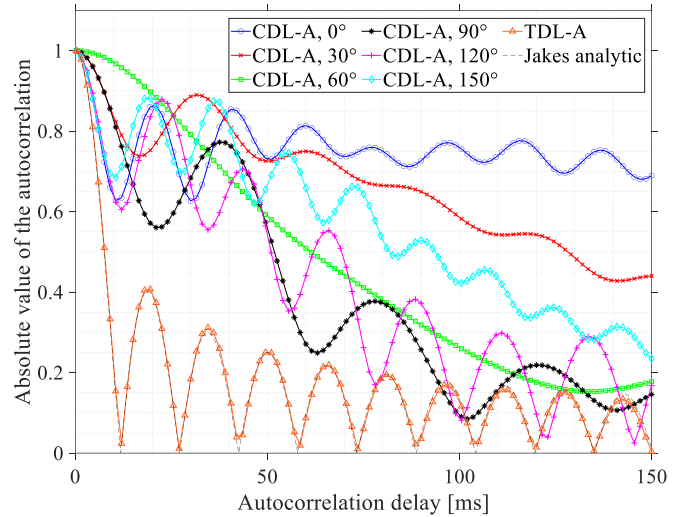

**Figure 1: The absolute value of the autocorrelation as a function of the autocorrelation delay for different channels.**

## II. The Time-domain Channel Property Feedback

In this section, we describe the autocorrelation metric whose amplitude and optionally also phase are carried by the TDCP report. Furthermore, we also give a brief discussion on its relation to Doppler spread.

### A. Autocorrelation Metric

The channel autocorrelation is a direct measure of how fast the channel varies with time and is thus a suitable measure for the use-cases we are considering. For the tapped delay line (TDL) channel models where each of the channel tap adheres to the Jakes model [6], the autocorrelation has the well-known form of the zeroth order Bessel function. For more realistic spatial channel models, like the clustered delay line (CDL) channel models [6], the form of the autocorrelation is, however, very different and can depend strongly on the direction of the UE. For the use-cases at hand, we are primarily interested in the behavior of the autocorrelation for small autocorrelation delays for which the autocorrelation is close to one.

To get a practical measure, we define the amplitude of the normalized instantaneous autocorrelation function as

$$c(t, \Delta t) = \frac{|\sum_{n=0}^{N-1} h_n(t + \Delta t) \cdot h_n^*(t)|}{\sqrt{\sum_{n=0}^{N-1} |h_n(t + \Delta t)|^2} \sqrt{\sum_{n=0}^{N-1} |h_n(t)|^2}}, \quad (1)$$

where $h_n(t)$ is the channel for subcarrier $n$ at time $t$. Note that for normalization we use in the denominator the geometric average over the two time-instances $t$ and $t + \Delta t$ of the zero-delay autocorrelation function in order to make the metric robust against automatic gain control (AGC). Also note that to get a metric that is robust also towards phase discontinuities e.g., due to frequency adjustments, we use the absolute value of the instantaneous autocorrelation function and refer to this quantity as the channel correlation amplitude. Figure 1 shows the absolute value of the autocorrelation as a function of the autocorrelation delay for different channels and UE directions for the same UE speed of 10km/h. As noted earlier, the autocorrelation differs significantly for different UE directions.

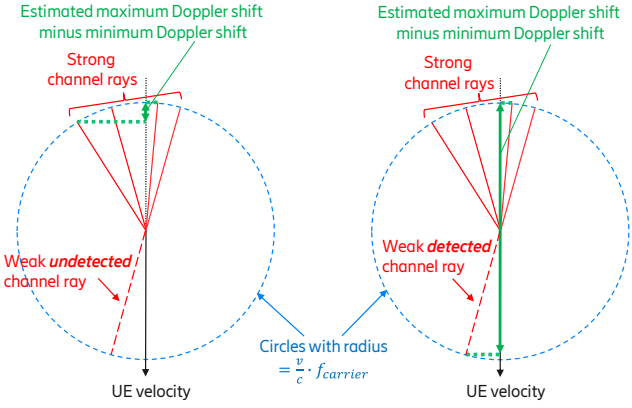

**Figure 2 : Doppler spread defined as the maximum Doppler shift minus the minimum Doppler shift is a bad measure of channel variability. A very weak channel ray could have an immense impact on Doppler spread while having negligible impact on channel variability. This measure would also be very volatile, depending on whether weak channel rays are detected above noise or not.**

Note that averaging or filtering over time should be done based on the absolute value to maintain robustness against phase discontinuities. This gives very small deviations compared to the ideal autocorrelation when the autocorrelation is close to one, which is the region where the low speeds can be distinguished. This metric is simple for a UE to estimate with respect to the conventional methods used for channel estimation.

The TDCP report can also optionally include the channel correlation phase. To estimate the correlation phase, the UE must know the network node's transmission frequency with very high precision. However, neither the precision of the UE oscillators nor the network node's oscillators are good enough to allow such knowledge. The best the UE can do is therefore to tune its frequency to the receive frequency, i.e. to perform frequency offset compensation. Using the receive frequency rather than the transmission frequency results in a linear phase rotation of the channel correlation estimate. As a consequence, the information on average Doppler shift, otherwise captured by the correlation phase, is lost.

### B. Relation to Doppler Spread

An alternative to reporting the absolute value of the autocorrelation function would be to report the Doppler spread as defined by the square root of the second moment of the Doppler power spectrum. The second moment of the Doppler power spectrum is proportional to the second derivative of the normalized autocorrelation function at zero delay [5] and the Doppler spread can thus be calculated based on the second order Maclaurin expansion of the normalized autocorrelation function as

$$f_d = \frac{\sqrt{-c''(0)}}{2\pi} \approx \frac{\sqrt{1-c(\Delta t)}}{\sqrt{2}\cdot\pi\cdot\Delta t} \quad (2)$$

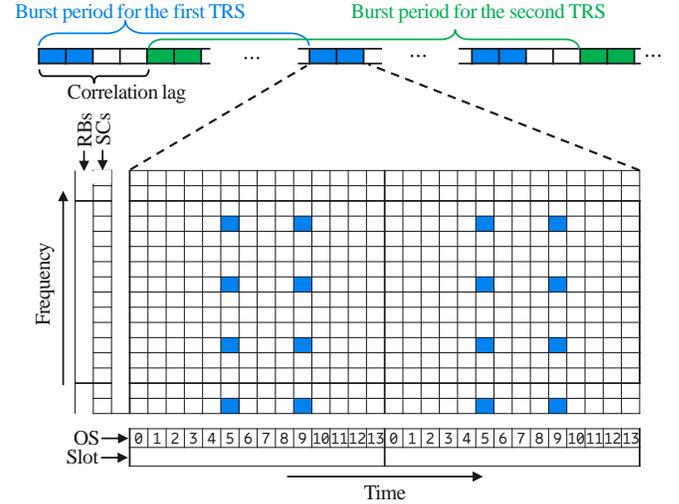

**Figure 3: To support estimation of the channel correlation over a delay of 4 OFDM symbols or 1 slot, it's sufficient to configure a single dual slot TRS, as illustrated by the blue TRS. To support a longer correlation delay it is, however, necessary to configure two TRSs with a relative time offset corresponding to the wanted correlation delay, as illustrated by the blue and green TRSs.**

when the delay $\Delta t$ is small. Doppler spread, as defined by the square root of the second moment of the Doppler power spectrum, thus gives good information about channel variability for low delays. The behavior of the autocorrelation function for large delays can, however, not be deduced from this measure.

Using the Bessel function form of the autocorrelation function for the Jakes model to define the Doppler spread is not suitable since realistic channels deviate strongly from the Jakes model as seen in Figure 1. Still, for low delays it would be equivalent to the method described above since the Maclaurin expansion of course is valid also for the Jakes model.

Another alternative would be to define Doppler spread as the maximum Doppler shift minus the minimum Doppler shift. This is, however, a bad measure of channel variability. A very weak channel ray has negligible impact on channel variability, but it could still have an immense impact on Doppler spread as illustrated in Figure 2. This measure would also be very volatile, depending on whether weak channel rays are detected above noise or not. This measure would also be very complex for the UE to estimate. Due to these reasons, it was agreed in 3GPP Release 18 to use autocorrelation metric as the reporting quantity for TDCP instead of Doppler Spread.

### C. Reporting format

The delay $\Delta t$ for which the UE should estimate and report the channel correlation is configured by the network. To measure the correlation over a large delay is complex for the UE since it needs the raw measurements of the received signal to be stored over the duration of the large delay. How large the configured delays can be therefore depends on the UE capabilities. For the most capable UEs the delay values 4 OFDM symbols (OS), 1 slot, 2 slots, 3 slots, 4 slots, 5 slots, 6 slots and 10 slots are supported while the least capable UEs only support delays up to 1 slot.

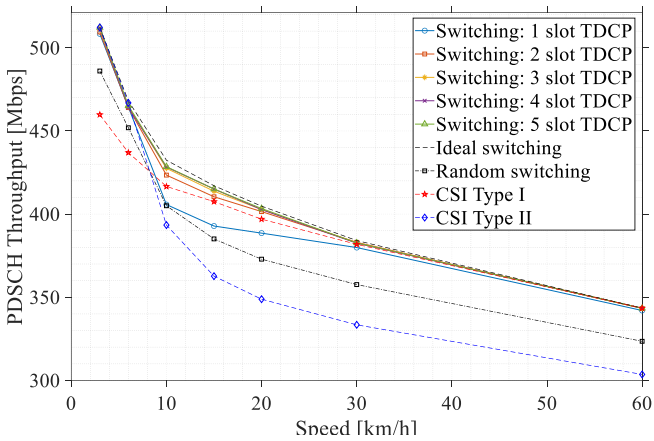

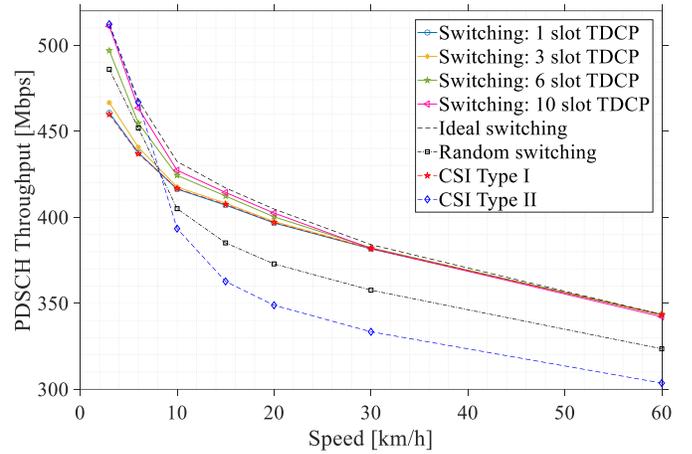

**Figure 4 Throughput for switching between CSI Type-I and CSI Type-II feedback-based precoding based on the estimated channel correlation amplitude over different correlation delays when TRS and PDSCH SNR both equal to 10dB. The cross over point between CSI Type I and CSI type II feedback is at roughly 8km/h. TDCP based switching based on a delay of 3 or more slots gives close to ideal switching performance. TDCP based switching based on a correlation delay of 1 slot gives rather bad performance while a delay of 2 slots gives decent switching performance.**

The correlation over small delays can be accurately estimated when the channel varies fast. However, when the channel varies slowly, then the correlation needs to be measured over a larger delay to achieve sufficient accuracy.

Depending on its capabilities, a UE may support simultaneous estimation and reporting of the correlation for up to 4 different delay values. The UEs may also have the additional capability to report not only the absolute value of the channel correlation but also the phase of the channel correlation.

## III. Configuration for TDCP

To estimate the channel correlation over a delay of 4 OFDM symbols or 1 slot, it is sufficient to configure a single dual slot TRS. In order to support a longer correlation delay it is, however, necessary to configure two TRSs with a relative time offset corresponding to the wanted correlation delay (see Figure 3). One of the TRSs would need to be configured with a rather high periodicity for the purpose of frequency and time tracking. The second TRS could however only be used for TDCP estimation and could be configured with a periodicity that is a multiple of the periodicity of the first TRS to reduce the overhead.

One typical configuration is a TRS burst of 2 TRS symbols in 2 adjacent slots repeated e.g., every 20ms.

In Release 18, TDCP is configured using the CSI reporting configuration framework. In each reporting configuration for TDCP, the number of delay values for which normalized autocorrelation amplitudes are reported are configured. The TRSs used for TDCP measurements are also configured as part of the reporting configuration for TDCP. Furthermore, whether the optional normalized autocorrelation phases for the configured number of delay values are reported or not can be controlled via configuration by the network node. In Release 18, the TDCP reporting is triggered aperiodically (i.e., using a similar mechanism as triggering aperiodic CSI).

**Figure 5 Throughput for switching between CSI Type I and CSI Type II feedback-based precoding based on the estimated channel correlation amplitude over different delays when TRS SNR equal to -2dB and PDSCH SNR equal to 10dB as can be the case if the TRS is scheduled to collide with TRSs from other cells rather than with PDSCH from other cells. In this case a correlation delay of 10 slots is needed to achieve good switching performance.**

## IV. Results

In this section, we show the benefit of TDCP report by evaluating two use-cases: a) Switching between CSI Type-I and Type-II precoding and b) DMRS time-direction density configuration.

### A. Evaluation Assumptions

The evaluation was performed based on link level simulations. The carrier frequency used was 3.5GHz. The subcarrier spacing was 30kHz and the slot length was 0.5ms. The system bandwidth as well as the TRS bandwidth was 100MHz. The CDL-A channel model [6] was used with 100ns delay spread, 45 degrees Azimuth angle of Arrival Spread (ASA) and 10 degrees Zenith angle of arrival spread (ZSA). The UE velocity was 3, 10, 20, 30 or 60km/h for the CSI Type-I Type-II switching use-case and 90, 120, 180, 250, 300, 350 or 500km/h for the DMRS density switching use-case. The UE had two antennas with different polarizations. The base station had an antenna array with two horizontal rows, four vertical columns and 2 polarizations. The horizontal antenna element spacing was half a wavelength and the vertical antenna element spacing was 0.8 times the wavelength. The antenna modelling followed the description in [6]. The periodicity for the CSI-RS as well as for the CSI Type-I and Type-II feedback was 20 slots. Adaptive coding, modulation and rank was used.

For the CSI Type-I/Type-II switching use-case, DMRS Type-I configuration with 1+1 additional DMRS symbols was used. For the DMRS time-direction density selection use-case, CSI Type-I and DMRS Type-I configuration with 1+1 additional or 1+2 additional DMRS symbols was used.

### B. Numerical Results

We first evaluate use-case 1, switching between CSI Type-I and CSI Type-II feedback. In Figure 5 we see that CSI Type-II feedback gives better performance than CSI Type-I at low UE speeds while CSI Type-I feedback gives better performance than CSI Type-II feedback at high UE speeds. The cross over point is at roughly 8km/h. The figure also shows the performance of a CSI Type I/Type II switching scheme based on a threshold on the estimated TDCP metric for different correlation delays. The TDCP metric used was the absolute value of the instantaneous autocorrelation function at a correlation delay of 1, 2, 3, 4 or 5 slots (0.5 to 2.5ms) as estimated based on two TRS bursts separated by the corresponding delay. The performance of TDCP based switching is close to the ideal performance for a correlation delay of 3, 4 or 5 slots. One may also note that the switching performance is better than ideal speed-based switching at 8km/h. This is due to the fact that the TDCP metric takes into account that the channel variability depends not only on UE speed but also on the UE direction and the spatial structure of the channel. In these results, Type I/Type II switching is performed for each UE direction and the throughput after switching is averaged over all the directions for a UE speed.

In Figure 5 we can observe that switching performance is highly dependent on what delay is used. If a small delay is used, then the channel variation over the delay is very small and hard to detect over noise. If on the other hand, a very large threshold is used then the corresponding threshold may be so low that the oscillating nature of the autocorrelation function creates ambiguities. We also see that a delay of two slots shows gains compared to random mode switching, but to get really good performance a delay with at least 3 slots is needed.

If the TRS is configured to collide with TRSs from other cells rather than with PDSCH from other cells, then the signal to interference plus noise (SINR) of the TRS can differ from the SNR of the PDSCH. If the SINR of the TRS is worse than the SINR for the PDSCH, an even larger delay may be needed to achieve good performance. In Figure 4 we see that a correlation delay of 10 slots is needed to get really good switching performance when the TRS SNR is -2dB and the PDSCH SNR is 10dB.

We now evaluate the second use-case, i.e., DMRS time-direction density configuration. Here, we limit ourselves to switching between 1 and 2 additional DMRS symbols in a slot, equivalent to 2 and 3 DMRS symbols in a slot, respectively. In Figure 6 we see that 1 additional DMRS symbol gives higher throughput at low UE speeds while 2 additional DMRS symbol gives higher throughput at high UE speeds. For 18dB SNR, the cross over point is at roughly 280km/h. Since the switching point is at such a high UE speed, we use a short autocorrelation delay of 4 symbols (0.143ms). Since the channel-variability is so high this low delay gives sufficient accuracy while avoiding ambiguities coming from the oscillations of the autocorrelation function at larger delays (see Figure 1). In Figure 6 we can see that the performance for TDCP based switching is close to ideal. Note that in this case the TDCP estimate is based on a single TRS burst.

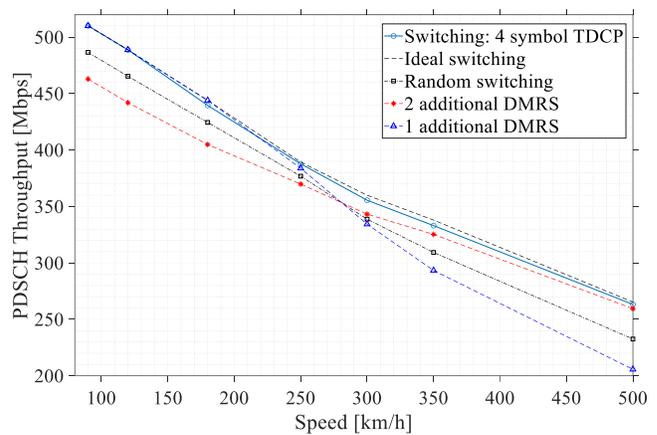

**Figure 6** Throughput for switching between 1 and 2 additional DMRS symbols based on the estimated channel correlation amplitude over a delay of 4 OFDM symbols at an SNR of 18dB. The cross over point between one and two additional DMRS symbols is at roughly 280km/h. TDCP based switching gives close to ideal switching performance.

We also evaluated for an SNR of 10dB and results showed that the switching point between 1 and 2 additional DMRS symbols is at a higher UE speed of roughly 400km/h. One should note that the switching point depends on the SNR and the TDCP threshold needs to be adjusted accordingly. The base station can do such an adjustment based on channel quality feedback or based on the selected modulation and coding scheme.

## V. CONCLUSION AND FUTURE DIRECTION

The channel correlation amplitude feedback gives significant throughput gain for the use-case of switching between CSI Type-I and Type-II based precoding as well as for switching between 1 and 2 additional DMRS symbols.

For switching between CSI Type-I and Type-II feedback, the switching point is at a low UE speed of roughly 8km/h. As a consequence, the correlation delay needs to be quite large in order to give sufficient accuracy for the TDCP measure. For the case where the TRS is scheduled to collide with PDSCH in neighboring cells, a correlation delay of at least 2 slots is needed. For the case where the TRS is scheduled to collide with TRS in neighboring cells, a correlation delay of at least 5 slots is needed.

For switching between 1 and 2 additional DMRS symbols, the switching point occurs at much higher UE speeds of roughly 250 to 400km/h, depending on the SNR. As a consequence, a short correlation delay is needed to avoid ambiguities coming from the oscillations of the autocorrelation function. A correlation delay of 4 symbols works well and allows TDCP measurements to be performed based on a single TRS burst.

The channel correlation amplitude feedback can be expected to be useful for a multitude of use-cases allowing the base station to optimize various configurations for optimal performance. Each use-case can be expected to require switching or decision points at different UE speeds. To cater for all use-cases the correlation delay can be flexibly configured between 4 symbols and 10 slots.

Since Release 18 TDCP reporting is performed after radio resource control (RRC) configuration when the UE is in RRC connected mode, decisions based on TDCP reports at the network side can only be made after the network receives a TDCP report after UE moves to RRC connected mode and after RRC configuration. This means that the network should first RRC configure the UE with TDCP reporting after the UE is in RRC connected mode, then receive the configured TDCP report from the UE, and thereafter do other configurations based on the TDCP feedback. This may result in large delays as the decisions on configurations that depend on the TDCP feedback can only be made after receiving the TDCP report. Hence, how to reduce TDCP acquisition delay is an open problem to be solved. To address this problem, one future avenue to explore is to enable early TDCP reporting where the UE feeds back TDCP report as part of random-access procedure.

### Acknowledgment

The authors would like to thank their colleagues at Ericsson who have been part of developing the tools used for producing the numerical results in this article.

**Per Ernström** (per.ernstrom@stosn.com) is a principal researcher at Ericsson AB working on the 5G/NR radio communication technology. He was manager for the Ericsson 3GPP RAN standardization program 2010–2016. He received his M.S. from KTH royal Institute of technology in 1989 and his Ph.D. in theoretical particle physics from Stockholm University 1994 and had a research fellowship at the Nordic center for theoretical physics at the Niels Bohr Institute in Copenhagen 1994 to 1996.

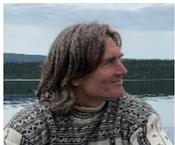

**Siva D. Muruganathan** (siva.muruganathan@ericsson.com) received his Ph.D. degree in electrical engineering from the University of Calgary, Canada, in 2008. He is currently a researcher at Ericsson Canada and a 3GPP RAN1 delegate working on 5G and 5G-Advanced standardization. He previously held research/postdoctoral positions at BlackBerry Limited, CRC Canada, and the University of Alberta, Canada. He was a Rapporteur for the 3GPP Release 15 study on LTE Aerials. He received the Inventor of the Year award from Ericsson in 2022, and the Fred W. Ellersick Prize from the IEEE Communications Society in 2021.

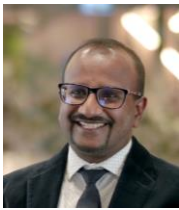

**Keerthi Kumar Nagalapur** (keerthi.kumar.nagalapur@ericsson.com) is a Senior Researcher at Ericsson Research, Gothenburg, Sweden. His current research activities focus on multi-antenna techniques and high-speed train communications. He received his M.Sc. and Ph.D. degrees in Electrical Engineering from Chalmers University of Technology, Gothenburg, Sweden in 2012 and 2018, respectively.

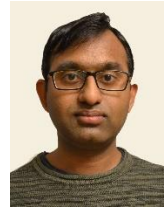

**Fredrik Athley** (fredrik.athley@ericsson.com) is a Master Researcher at Ericsson Research, Gothenburg, Sweden working on multi-antenna techniques and systems. He received his M.Sc. and Ph.D. degrees in electrical engineering from Chalmers University of Technology, Gothenburg, Sweden in 1993 and 2003, respectively. He also received the International Diploma of Imperial College, London, UK in 1993.

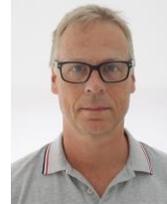